\documentclass[english,letter]{ieej-e}
\usepackage{graphicx}
\usepackage{subfigure}
\usepackage[varg]{txfonts}

\FIELD{C}
\YEAR{2021}
\NO{2}
\title[RDH with Digital Halftoning for Printing Technology]{Reversible Data Hiding Associated with Digital Halftoning That Allows Printing with Special Color Ink by Using Single Color Layer}
\authorlist{%
 \authorentry{Minagi Ueda}{n}{CUI}
 \authorentry{Shoko Imaizumi}{m}{CUI2}
}
\affiliate[CUI]{Graduate School of Science and Engineering, Chiba University,\\1-33 Yayoicho, Inage-ku, Chiba-shi, Chiba, Japan 263-8522}
\affiliate[CUI2]{Graduate School of Engineering, Chiba University,\\1-33 Yayoicho, Inage-ku, Chiba-shi, Chiba, Japan 263-8522}

\received{2020}{4}{9}
\revised{2020}{6}{25}

\begin{document}
\begin{abstract}
We propose an efficient framework of reversible data hiding to preserve compatibility between normal printing and printing with a special color ink by using a single common image. The special color layer is converted to a binary image by digital halftoning and losslessly compressed using JBIG2. Then, the compressed information of the binarized special color layer is reversibly embedded into the general color layer without significant distortion. Our experimental results show the availability of the proposed method in terms of the marked image quality.
\end{abstract}
\maketitle

\section{Introduction}
Copyright protection of digital content is important due to the spread of Internet services and increasing opportunities for uploading and sharing images. Data hiding techniques to prevent unauthorized copying and secondary use by embedding the copyright information into the image have been actively studied. Data hiding techniques can be divided into two types: irreversible data hiding (IDH) and reversible data hiding (RDH) \cite{RDH20y,HS,DE,PEEHS}. The former is resilient against some attacks and has high hiding capacity. However, the quality of the retrieved image slightly deteriorates after the payload is extracted. In contrast, the latter is usually more vulnerable to such attacks and has less hiding capacity than IDH, but it can completely retrieve the original image.

Normal printing is based on subtractive color mixing using cyan (C), magenta (M), yellow (Y), and black (K) inks. Image information for normal printing originally consists of RGB components, and it is converted to CMYK color space on printing. Due to the development of printing technology, it has been possible to represent textures and special colors, such as gloss, fluorescence, and white, which are difficult to express using only CMYK inks. To print a special color, it is necessary to add an extra layer to the general RGB color layer.

In this paper, we propose an efficient framework for preserving compatibility between normal printing and printing with a special color ink by using a single common image. Here, we assume that the target image is a natural image and the special color ink is a silver ink. The layer with the silver ink, the silver layer, is converted to a binary image by digital halftoning and compressed to appear to have multiple tones with less information. Then, the silver layer is hidden in the general color layer without visible artifacts by using an RDH method. Accordingly, we can suppress the total information amount of the color layers by integrating multiple layers into a single one.

\section{Proposed method}
We propose a new framework for printing with a special color ink. We prepare an 8-bit silver layer, which is arbitrarily designed based on the general color layer according to the user's request. The structure of the silver layer determines the area and density of the silver ink and is converted to the corresponding binary image using a dither method \cite{Dither} to reduce the information amount without sacrificing the representation of multiple tones. Each 8-bit pixel is changed to a binary pixel by using a dither matrix. On printing, an input image is binarized by digital halftoning using a printer driver. We have confirmed that major printer manufacturers have introduced digital halftoning using a dither method. Thus, the prior binarization of the silver layer using a dither method does not seriously degrade the quality of the printed matter in the proposed method. Then, the binarized silver layer is compressed with JBIG2 \cite{JBIG2}, which is an international image compression standard for binary images. The JBIG2 codec has both lossy and lossless compression options. We use lossless compression in the proposed method to perfectly retrieve the binarized silver layer. 

The compressed layer information is embedded into the general color layer by RDH. Our embedding algorithm is based on the prediction error expansion and histogram shifting (PEE-HS) method \cite{PEEHS}. The PEE-HS method calculates the prediction value $u'_{m,n}$ for each pixel $u_{m,n}$ using four neighboring pixels. $u'_{m,n}$ is given as 
\begin{equation}
\label{eq1}
u'_{m,n} = \left \lfloor \frac{u_{m,n-1}+u_{m+1,n}+u_{m,n+1}+u_{m-1,n}}{4} \right \rfloor .\\
\end{equation}
The prediction error $d_{m,n}$ is obtained by
\begin{equation}
\label{eq2}
d_{m,n} = u_{m,n} - u'_{m,n}.\\
\end{equation}
The payload $w$ is embedded into the prediction error $d_{m,n}$ according to the following equation.
\begin{eqnarray}
	\label{eq3}
	D_{m,n}=\left\{ \begin{array}{ll}
	2d_{m,n} + w, & \textrm{if} \  d_{m,n} \in [-T,T)\\
	d_{m,n} + T, & \textrm{if} \  d_{m,n} \geqq T\\
	d_{m,n} - T, & \textrm{if} \  d_{m,n} < -T.\\
	\end{array} \right.
\end{eqnarray}
The threshold value $T$ is determined in accordance with the payload amount and is a positive value. Each marked pixel $U_{m,n}$ is attained by
\begin{equation}
	U_{m,n} = D_{m,n} + u'_{m,n}.\\
\end{equation}
The payload $w$ consists of not only the silver layer information but also additional information, e.g., a location map and threshold value $T$. 

In the extraction procedure, the prediction value is calculated in a similar fashion. After payload extraction, the silver layer is decompressed with JBIG2, and then the original information of the general color layer and binarized silver layer can be retrieved perfectly.

\begin{figure}[tb]
\begin{center}
	\subfigure[General color layer]{
		\includegraphics[width = 30 mm]{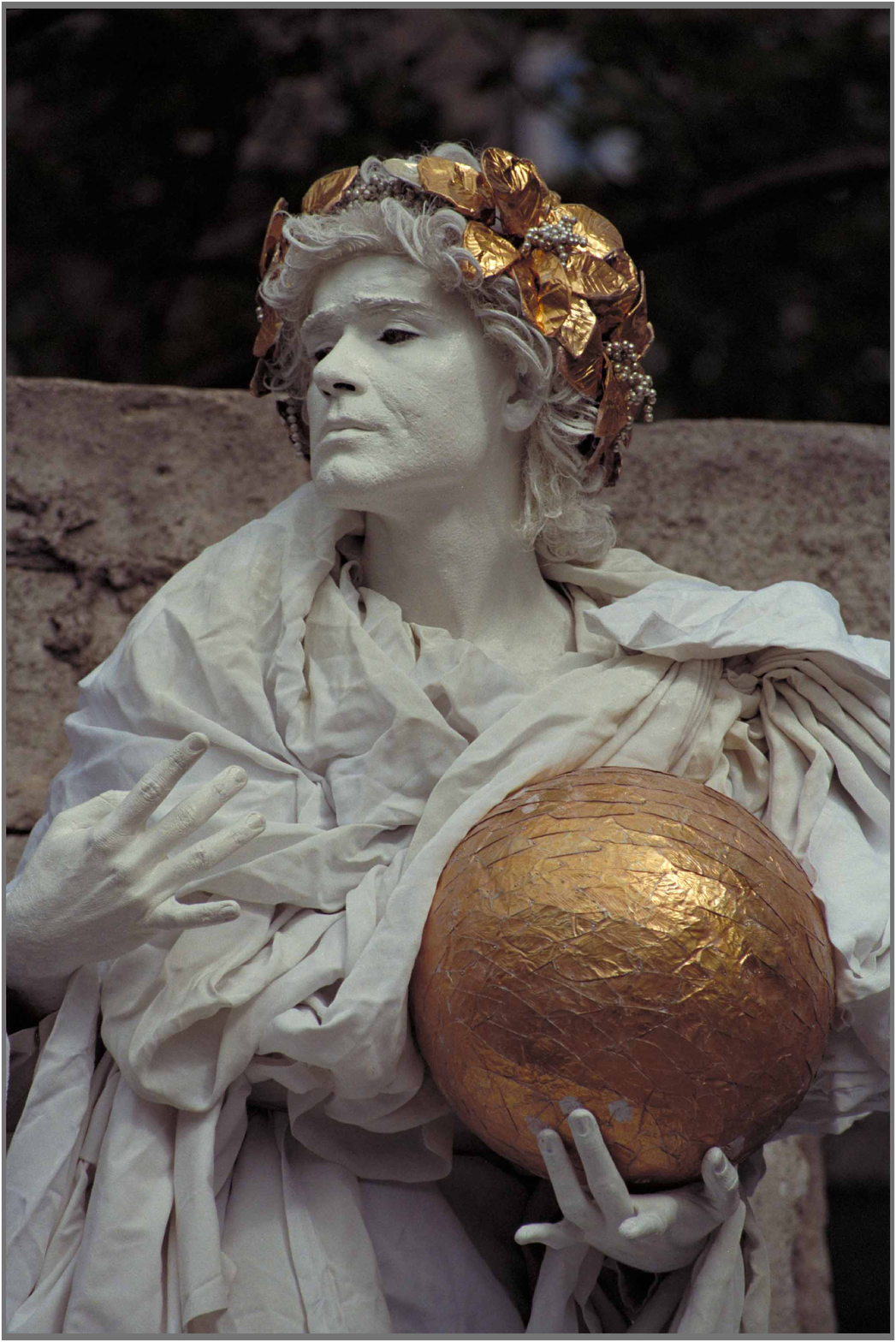}
		\label{Original_Color}
	}
	\subfigure[8-bit silver layer]{
		\includegraphics[width = 30 mm]{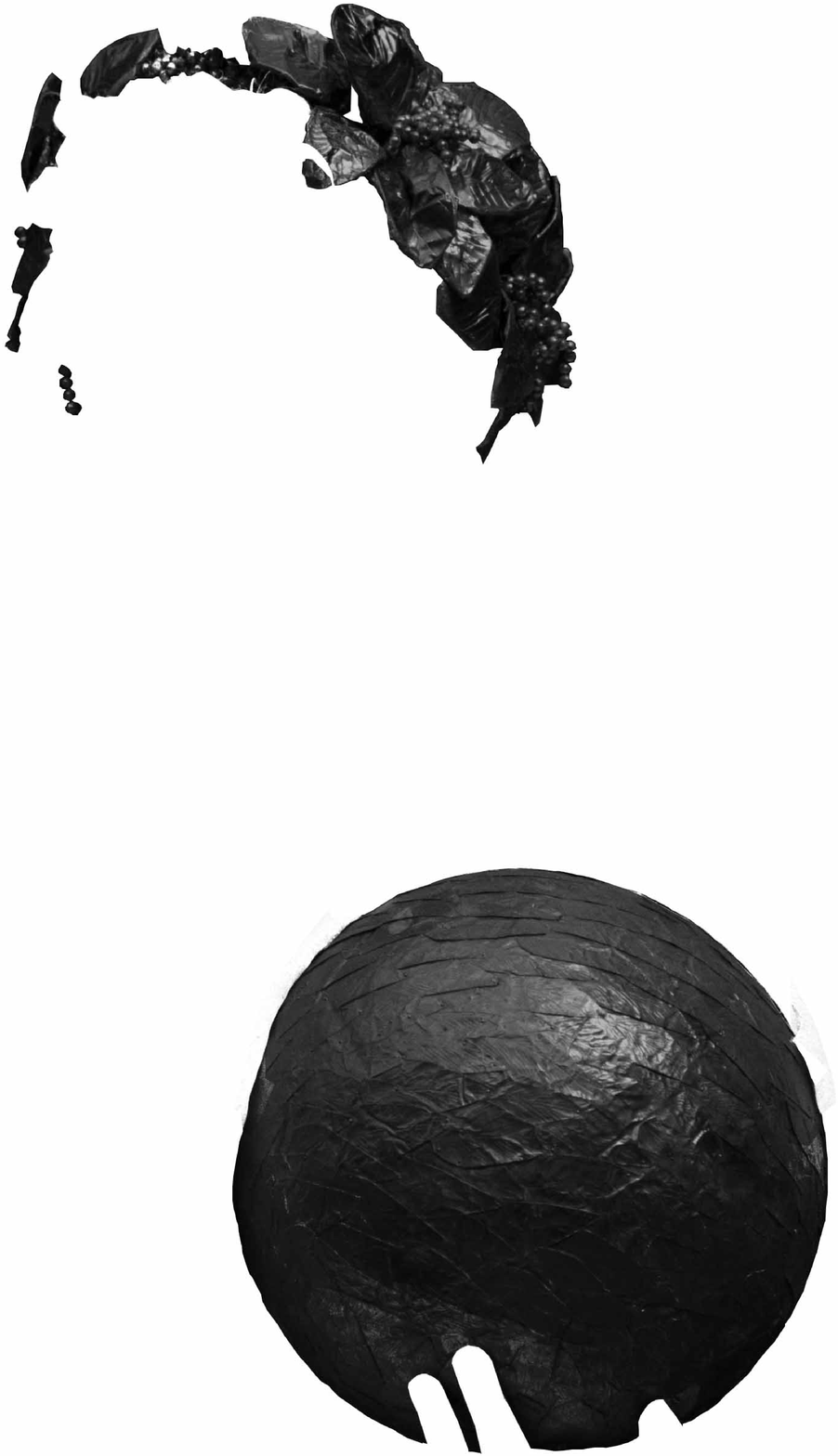}
		\label{Grayscale}
	}
\end{center}
\caption{Example of test images ($3,072 \times 2,048$ pixels).}
\label{Original}
\end{figure}

\begin{figure}[tb]
\begin{center}
	\subfigure[Bayer]{
		\includegraphics[width = 25 mm]{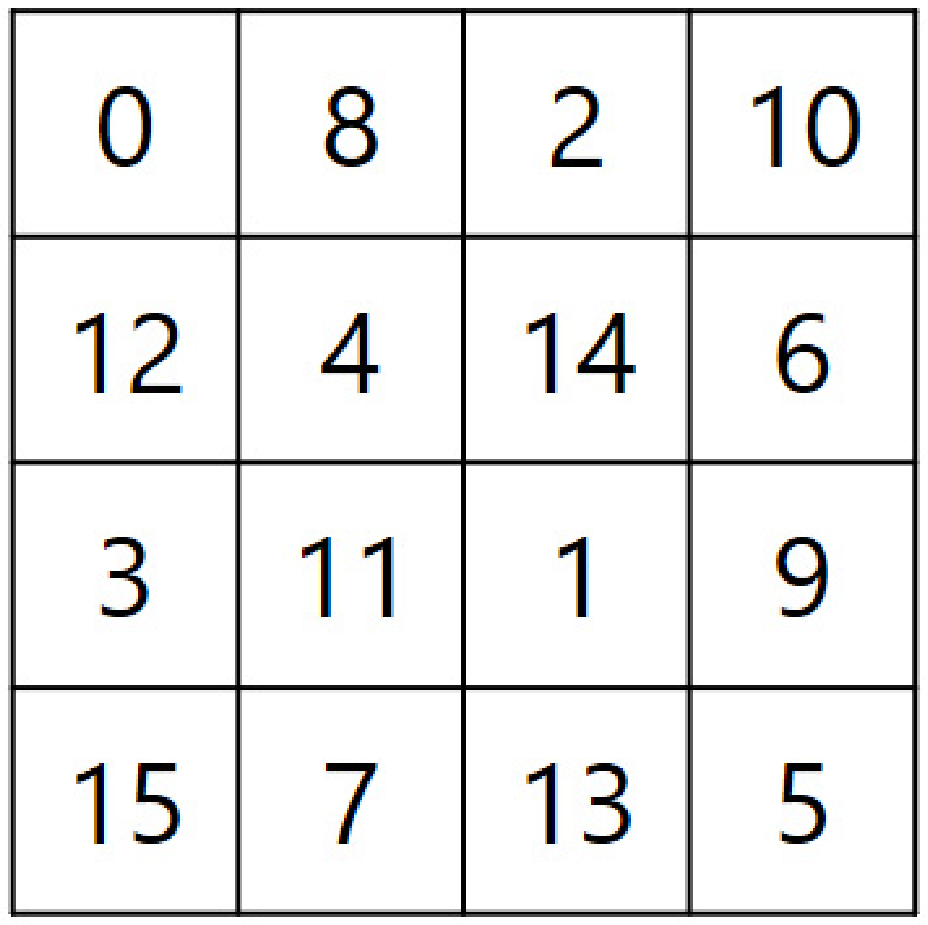}
		\label{Bayer_Matrix}
	}
	\subfigure[Halftone]{
		\includegraphics[width = 25 mm]{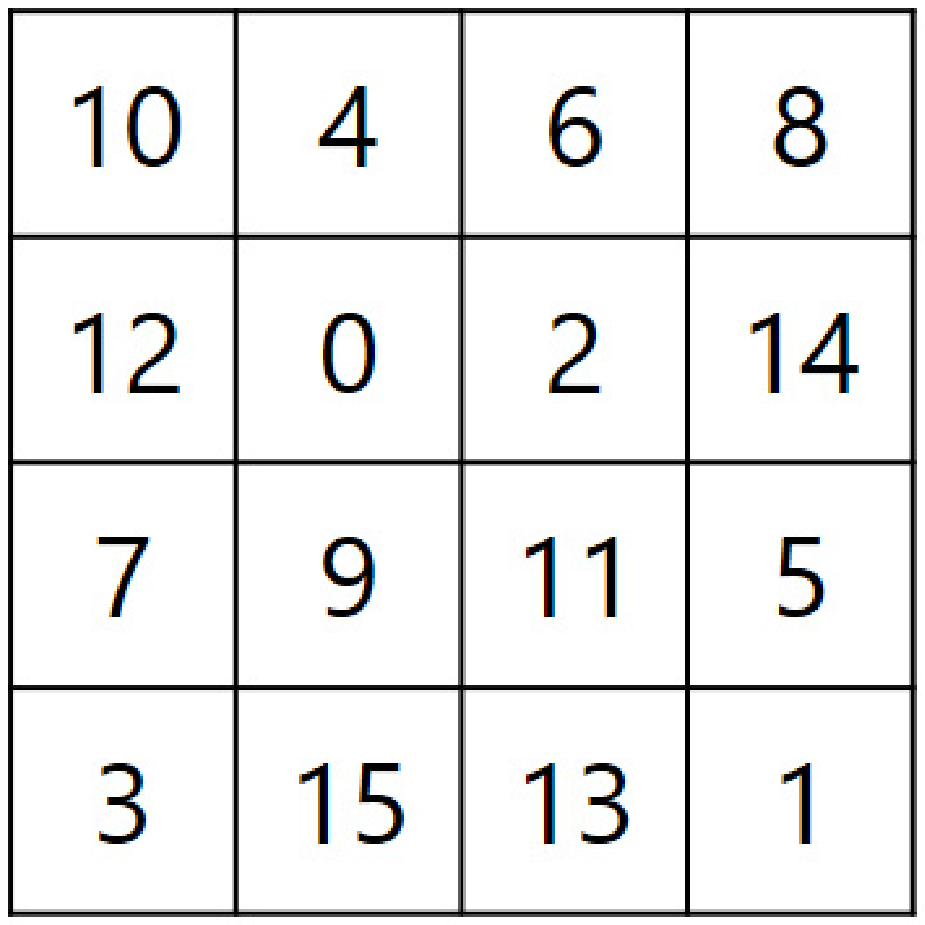}
		\label{Bayer_random_Matrix}
	}
\end{center}
\caption{Dither matrices.}
\label{Dither_Matrix}
\end{figure}

\section{Experimental Results}
We verify the effectiveness of the proposed method using five test images in our experiments. Figures \ref{Original_Color} and \subref{Grayscale}  show a general color layer, that is, an original image, and an 8-bit silver layer, respectively. The silver layer is first binarized by using a dither matrix, which is shown in Fig.\ref{Dither_Matrix}. Then, the silver layer is losslessly compressed with JBIG2. The compression ratio was nearly 90\% for each binarized silver layer. Subsequently, we embed the compressed layer information with additional information into a single color component of the general color layer. 

\begin{figure}[tb]
\begin{center}
	\subfigure[$\scriptstyle \mbox{Bayer}\atop \scriptstyle \mbox{(payload: 615,406 bits)}$]{
		\includegraphics[width = 30 mm]{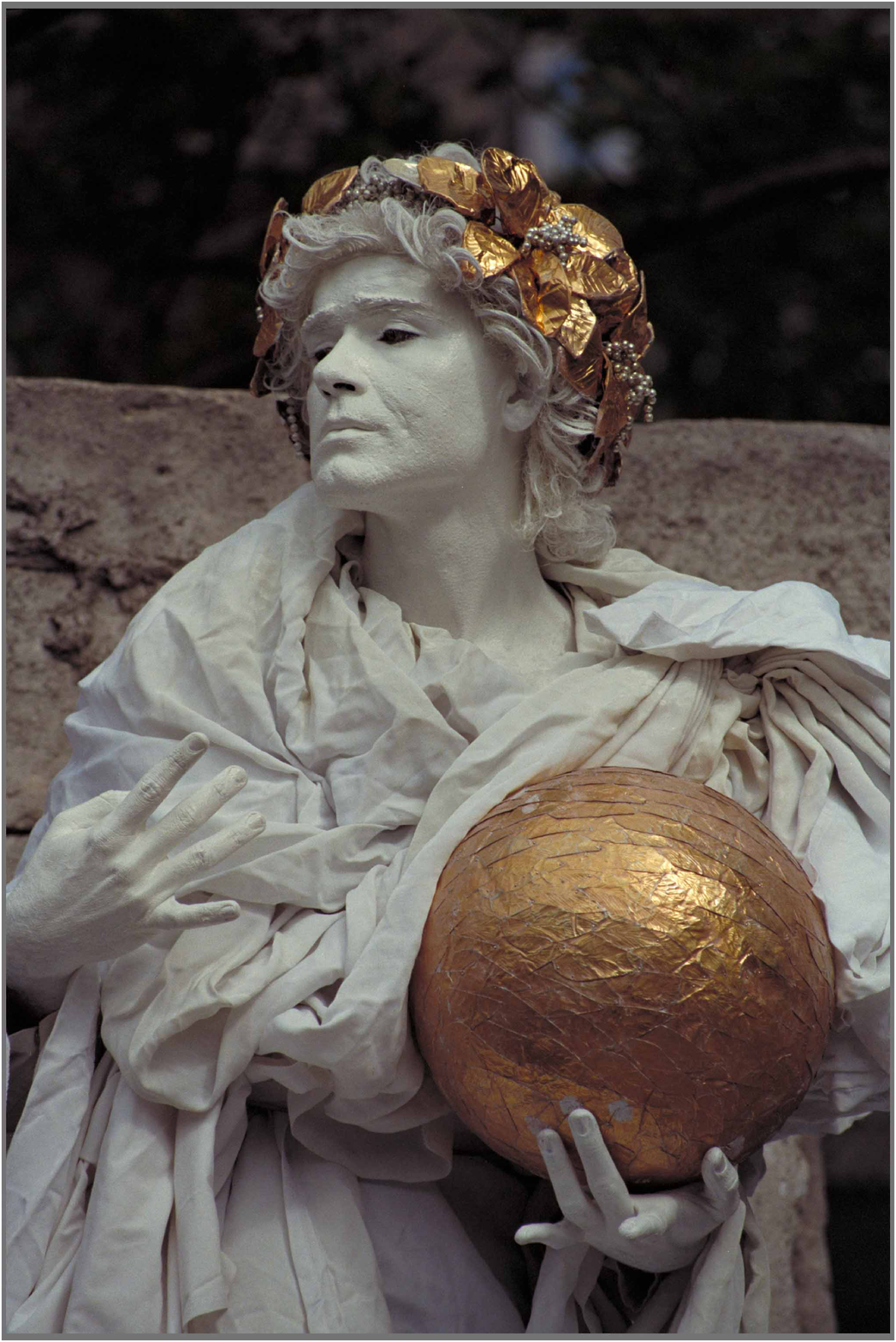}
		\label{Embed_Bayer}
	}
	\subfigure[$\scriptstyle \mbox{Halftone}\atop \scriptstyle \mbox{(payload: 609,478 bits)}$]{
		\includegraphics[width = 30 mm]{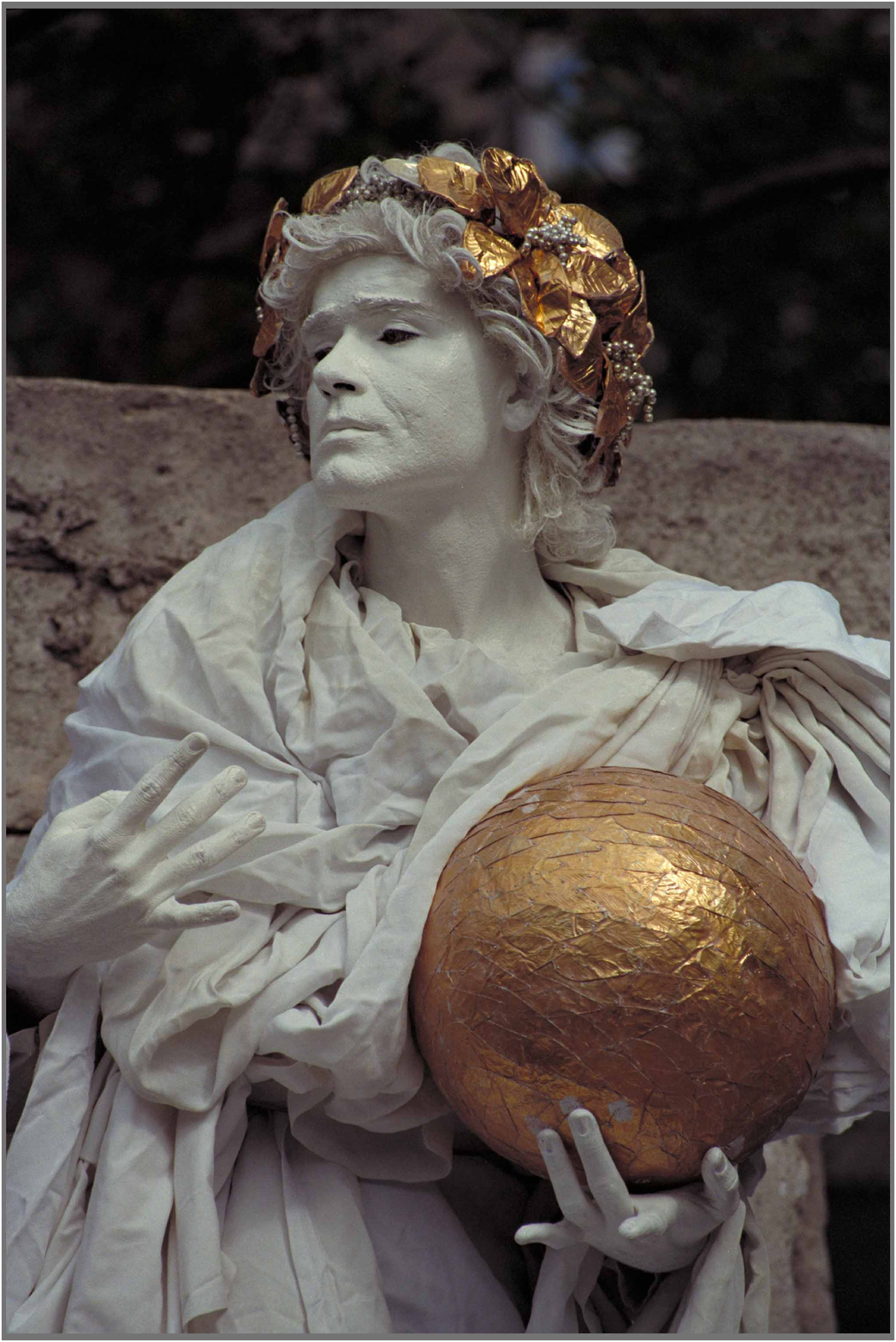}
		\label{Embed_Bayer_random}
	}
\end{center}
\caption{Marked images.}
\label{Marked_Image}
\end{figure}

Figure \ref{Marked_Image} depicts the marked images, where all the payload has been hidden in the B component. Table \ref{tb1} shows the payload amount and the PSNR and MSSIM values of the marked images. Each value indicates the mean of the five test image values. As the payload amount becomes smaller, the degradation in the quality of the marked image can be suppressed more efficiently. The information amount of the silver layers can be significantly reduced by both digital halftoning and JBIG2 compression. Furthermore, the silver layers are embedded into the general color layers using the PEE-HS method, which is an effective RDH method. Accordingly, we obtain the marked images without visible artifacts. When the payload is hidden in the B/G component, the PSNR values become highest/lowest in the three color components. This is because the human eye has a lower/higher sensitivity to the B/G component than the other color components. It is confirmed that the marked image quality can be kept high without severe distortion in all cases.

\begin{table}[tb]
\caption{Image quality evaluation.}
\begin{center}
	\begin{tabular}{|c||c|r|r|r|}
	\hline
	\textbf{Marked color component} & \textbf{Matrix} & \textbf{Payload [bpp]} & \textbf{PSNR [dB]} & \textbf{MSSIM} \\ \hline
	R & Bayer & 0.085 & 57.33 & 0.9990\\
	   & Halftone & 0.082 & 57.55 & 0.9990\\ \hline
	G & Bayer & 0.076 & 53.76 & 0.9975\\
	   & Halftone & 0.073 & 54.42 & 0.9981\\ \hline
	B & Bayer & 0.091 & 61.75 & 0.9996\\
	   & Halftone & 0.088 & 61.98 & 0.9996\\ \hline
	\end{tabular}
\label{tb1}
\end{center}
\end{table}

\section{Conclusion}
We proposed an effective reversible data hiding application to maintain compatibility between normal printing and printing with a special color ink using a single common image. In normal printing, the marked image can be directly printed with an ordinary printer. In contrast, when printing with a special color ink, the general and special color layers are restored from the marked image, and then both the layers are sent to the dedicated printer. We evaluated the quality of the marked images using PSNR and MSSIM. The experimental results proved that the marked image quality is quite high without visible artifacts.

\section*{Acknowledgments}
This work was partially supported by the Okawa Foundation for Information and Telecommunications and the Institute for Global Prominent Research, Chiba University.

\end{document}